# Characterization of Magnetic Ni Clusters on Graphene Scaffold after High Vacuum Annealing

*Zhenjun Zhang, Akitomo Matsubayashi, Benjamin Grisafe, Ji Ung Lee, James R. Lloyd

College of Nanoscience and Engineering, SUNY Polytechnic Institute, Albany, NY 12203

## Abstract

Magnetic Ni nanoclusters were synthesized by electron beam deposition utilizing CVD graphene as a scaffold. The subsequent clusters were subjected to high vacuum (5-8 $\times 10^{-7}$ torr) annealing between 300 and 600 $^0$C. The chemical stability, optical and morphological changes were characterized by X-ray photoemission microscopy, Raman spectroscopy, atomic force microscopy and magnetic measurement. Under ambient exposure, nickel nanoparticles was observed to be oxidized quickly, forming antiferromagnetic nickel oxide. Here, we report that the majority of the oxidized nickel is in non-stoichiometric form and can be reduced under high vacuum at temperature as low as 300 $^0$C. Importantly, the resulting annealed clusters are relatively stable and no further oxidation was detectable after three weeks of air exposure at room temperature.

## 1. INTRODUCTION

Synthesis of nanometer scale metal clusters has attracted tremendous attention in recent years due to their unique properties, compared to the bulk counterparts [1]. These properties enable metal clusters to have possible applications in the area of biomedical drug delivery, catalysis, imaging contrast enhancement and biochemical sensing [2-5]. Among all metal nanoclusters, magnetic metals such as nickel and cobalt are of particular interest. There is a wide range of reported methods to fabricate nickel or cobalt nanoclusters that can generally be classified in three different categories 1. Chemical reduction or decomposition 2. Physical vapor deposition. 3. Electrochemical deposition [6-10]. Graphene, a $sp^2$ hybridized carbon material, has already been widely studied for a variety of applications in electronics, spintronics and energy storage [11-13]. Hybridization of graphene with nanoclusters has shown potential in electrochemical devices and high performance catalysis [14, 15]. In addition, nanoclusters grown on graphene with low defect density can be used as thermal and electric carriers [16, 17]. Metal clusters on graphene sheets can be generated using wet chemistry but it has a tendency to sacrifice the conduction property of graphene. Due to lack of oxygen anchors, metals deposited on graphene tend to ball up and form clusters [18]. Using this property, it is possible to assemble metal decorated graphene on an arbitrary substrate.

To date, numerous reports have been focused on the study of well dispersed metal arrays on the CVD graphene moiré on close-packed metal surfaces. In these reports, high arrays of quality metal clusters such as (Re, W, Pt, Ir, Ni) can be synthesized in ultra-high vacuum using well controlled deposition techniques and highly ordered substrates [19-23]. However, fabrication of metal nanoclusters on transferred graphene with arbitrary substrates is not well studied. Furthermore, the stability of some of the oxophilic metals is not well characterized. Specifically, nickel thin films are vulnerable to oxidation at ambient condition and it is expected this effect can be worsened if the nickel is in the form of nanoclusters, in which large surface areas are in contact with the ambient environment.

When unprotected, nickel clusters are usually in a partially oxidized condition [24]. Partially oxidized nickel or cobalt nanoparticles are well known to exhibit hysteresis loops along the magnetic field axis, namely exchange bias (EB) [25, 26]. It happens when ferromagnetic components (nickel or cobalt) is exchange coupled with an antiferromagnetic component (nickel or cobalt oxide) at the interface. This property is useful not only in spintronic devices but also for probing the properties at the ferromagnetic-antiferromagnetic interface [27]. Very recently, cobalt oxide has been reported to be partially reduced at high vacuum annealing conditions and showed significant reduction of exchange bias afterwards [27].

Here we demonstrate a simple method to hybridize nickel nanoclusters with graphene on an arbitrary substrate, investigate the oxidation effects of electron beam deposited nickel nanoclusters on graphene under ambient exposure and study its change of properties after high vacuum annealing. Our XPS measurements show that the majority of the oxidized nickel is in nonstoichiometric form. We observed that nonstoichiometric nickel oxide can be reduced at temperatures as low as 300 $^0$C under high vacuum annealing conditions. At 600 $^0$C, nearly 95% of the oxide was transformed to the metallic form. Our magnetic measurements detect the hysteresis loop shift after ambient exposure but the exchange bias became symmetric after high temperature vacuum annealing, which is supplemental evidence for the reduction of nonstoichiometric nickel oxide.

## 2. EXPERIMENTAL

CVD graphene was used as the scaffold for the growth of nickel nanoclusters. The graphene was grown on a 25 micrometers thick commercially available polycrystalline Cu foil and was then dry transferred to a silicon substrate with 90 nm thermally grown $SiO_2$. To transfer graphene onto $SiO_2$, poly(methylmethacrylate) (PMMA) was coated onto as-grown CVD graphene on Cu foil and baked at 65 $^0C$ for 5 min to evaporate the solvent, followed by backside etching using a Technics 800 Micro RIE chamber with $O_2$ Plasma (30W, 30 Seconds). A thermal release tape was then attached to the PMMA film with a cut window in the center and floated on an ammonium persulfate solution (APS). After the copper residue was etched away and washed with de-ionized water, the graphene with PMMA support was transferred to a $SiO_2$ substrate. The PMMA layer was then removed in an acetone bath for 1 hr, followed by an isopropyl alcohol (IPA) rinse and dried in $N_2$ gas. The quality and layer of the as-transferred graphene sheet was monitored by Raman spectroscopy at 532 nm wavelength. The samples were then loaded into an electron beam evaporator chamber with a commercially available 99.9% pure nickel crucible and pumped down to a base pressure below $5 \times 10^{-7}$ Torr. The nickel crucible was then locally heated by electron beam to reach the vaporizing pressure and deposited at a rate of 1A/s for 30 second on two different substrates 1.nickel on graphene/$SiO_2$ 2.nickel on $SiO_2$. The samples are taken out of vacuum chamber and exposed to ambient air for 24h. The chemical state of the samples were characterized by both a Thermo Scientific Theta Probe XPS instrument and Verserlab Vibrating Sample Magnetometer (VSM). The morphology of the nickel nanoclusters was characterized with a Veeco Dimension 3100 AFM instrument.

After the initial characterization, the sample was loaded into a custom made vacuum chamber with a base pressure of $6 \times 10^{-8}$ torr. After a quick temperature ramp that ranged between 300 to 600 $^0$C, the temperature was maintained for 1 hr, at a vacuum pressure between 5-8 x$10^{-7}$ torr. The sample was then cooled down to room temperature by convection and further characterized with XPS and Verserlab VSM for comparison.

## 3. RESULTS AND DISCUSSION

Figure 1a and 1b shows the Ni 2p core emission XPS data after 24 hr air exposure and 1 hr high vacuum annealing (5x$10^{-7}$ torr) at 300 $^0$C. Because of the conductive nature of graphene, the sample under testing did not show a noticeable amount of charge shifting by measurement of the adventitious 1s carbon peak in the sample. In Figure 1a, the Ni 2p peak spectrum shows at least three distinct nickel species: (1) the metallic nickel (2) stoichiometric nickel oxide (3) $NiO_x.yH_2O$. Taking account the spin-orbit splitting effect, the fitted peak at 852.7 and 853.8 eV correspond to the metallic nickel and NiO $2p_{3/2}$ binding energy, respectively. Another Ni $2p_{3/2}$ peak at 852.87 eV is assigned to a generic $NiO_x.yH_2O$, due to the complication of multiplet splitting and plasmon loss effects for nickel 2p spectrum. To assure that the minor shift of the nickel 2p peak is not due to hybridization with graphene, a reference study was carried out on nickel deposited on $SiO_2$ and a similar result was obtained. Angle dependent XPS spectrum (not shown) confirmed that the nickel cluster is oxygen rich on the surface and nickel rich inside the core. The mechanism of oxidation effect on the as-deposited nickel nanoclusters is presumably related to its high surface to volume ratio and large oxygen concentration difference between that in equilibrium with a vacuum and ambient conditions. The as-deposited nickel clusters might have a high concentration of defects after e-beam evaporation. After ambient exposure, oxygen diffuses rapidly into the nickel cluster and occupies the defective sites to form a non-stoichiometric oxide. In other words, the formation of non-stoichiometric oxide is kinetically driven and it is in metastable state at room temperature. After 1 hour of high vacuum annealing, the XPS spectrum of the sample became cleaner with less background noise, which is an indication of less roughness at the surface. The nickel $2p_{3/2}$ spectrum is fitted with a main peak at 852.7 eV and two broad satellite peak (FWHM 2.5 eV) with binding energy 856.4eV (minor), 858.7 eV (major) [28]. The disappearance of a major peak at 853.8 is attributed to the reduction of majority of the $NiO_x$. Rationally, this unusual reduction could be attributed to either direct dissociation of nickel oxide to the vacuum or reduction of nickel oxide with underlying substance. The standard free energy for nickel oxide to directly dissociate into metallic nickel and oxygen molecule is 187.3 kJmol$^{-1}$, which is energetically unfavorable. On the other hand, the standard free energy for carbon species to reduce nickel oxide into metallic nickel and carbon dioxide is -20.6 kJmol$^{-1}$(300 $^0$C). The questions comes to whether it is the graphene associated carbon involved in reduction. In parallel experiment for nickel/$SiO_2$ showed similar results. It is reported that transition metal oxide can undergo reduction at temperature below 300 $^0$C in high vacuum condition ($2 \times 10^{-8}$ torr), assisted by surface carbon contaminants [29]. Therefore, small amount of surface carbon in our vacuum chamber can be the origin of reduction source. The accumulation of oxide at nickel surface and reduction was further evidenced by the 1s O peak in Figure 2. After ambient exposure, a significant amount of oxygen from nickel was detected. At 300 $^0$C in high vacuum annealing, the amount of oxide peak from nickel is dramatically reduced. After 600 $^0$C high vacuum anneal, the amount of oxygen in nickel clusters is further reduced to around 5% based on the normalized $NiO_x$ 1s O peak intensity ratio in the XPS spectra. Even though the absolute value cannot be judged, the relative amount of nickel oxide decreases to the minimum level at this temperature. Due to the fact that the whole measurement is performed ex situ, it cannot be ruled out that small amount oxidation may occur during the transfer process.

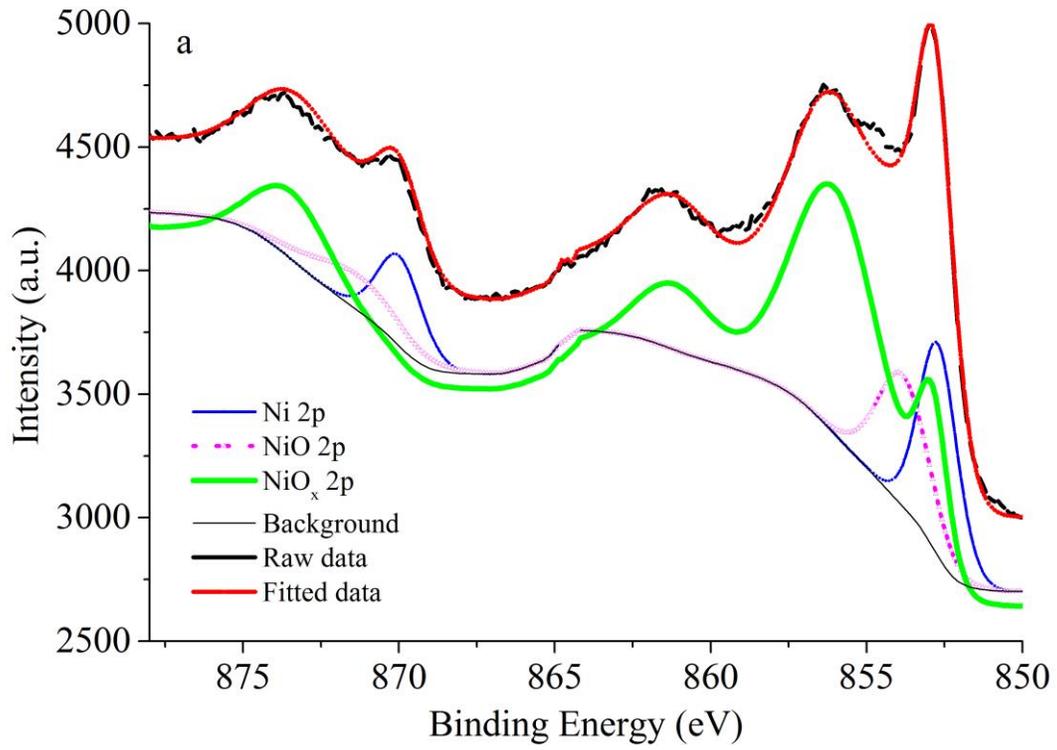

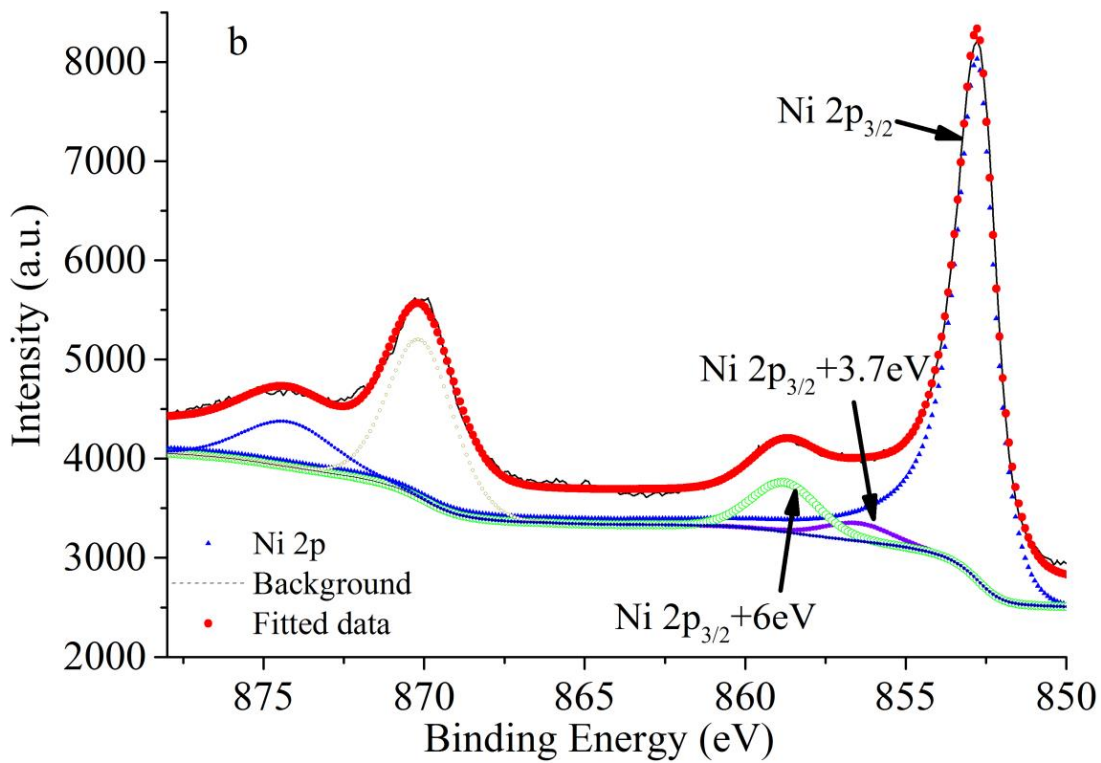

Figure 1. (Color online) Ni 2p X-ray photoelectron peaks: a) as-grown, b) 300 $^0$C annealed.

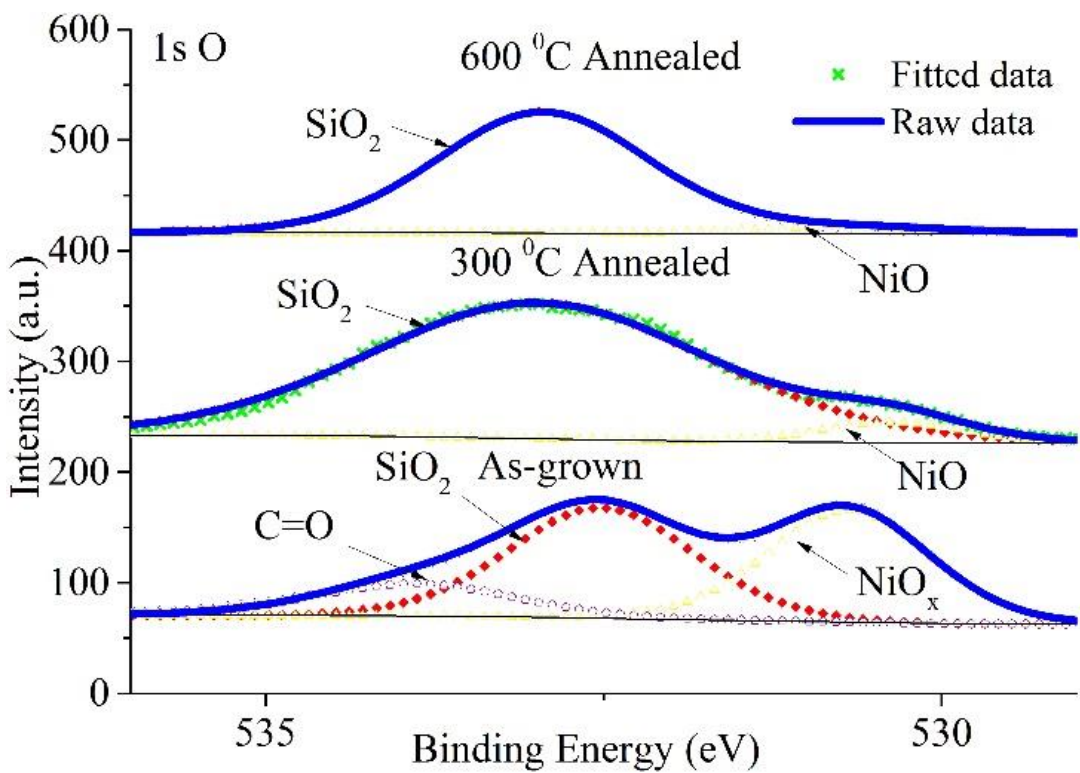

Figure 2. (Color online) Evolution of oxygen 1s X-ray photoelectron peaks: (1) as-grown, (2) $300^0$ C annealed, (3) 600 $^0$C annealed.

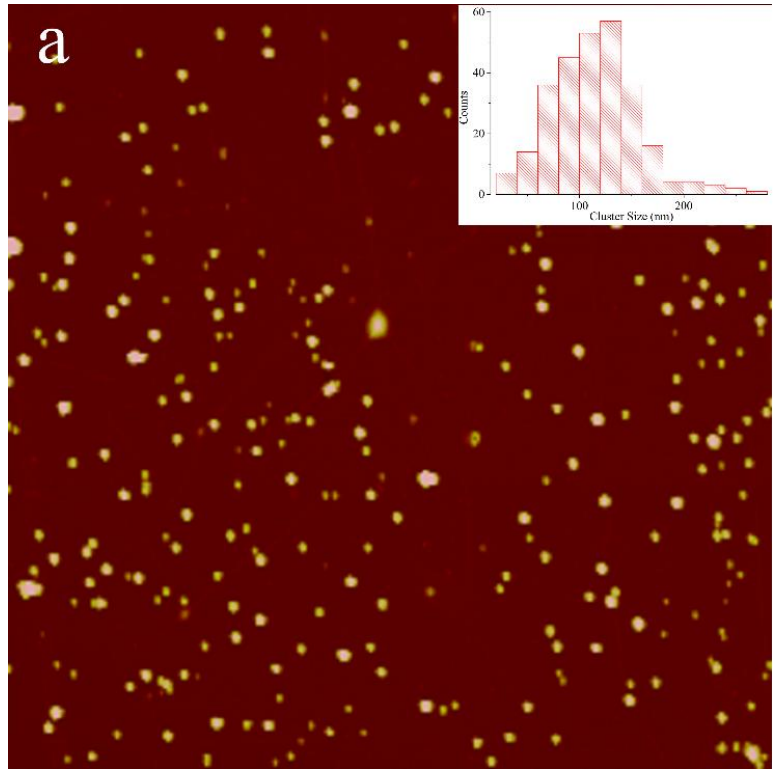

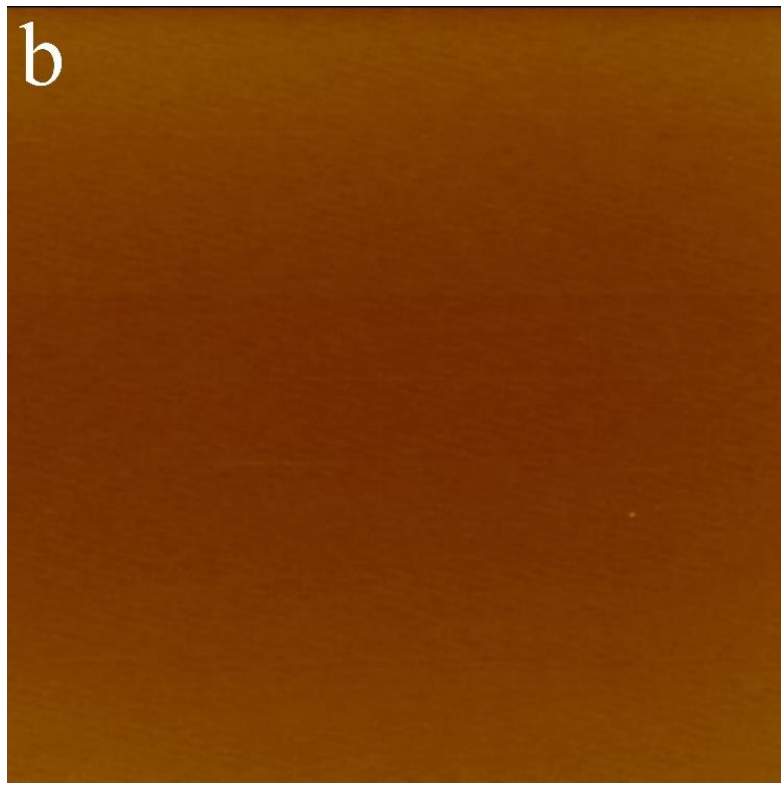

Figure 3. (Color online) AFM images (10x10 μm) of nickel sample on: a) graphene/$SiO_2$ with cluster size distribution on right side inset, b) $SiO_2$.

AFM was used to study the morphology of the nickel nanoclusters on graphene. Figure 3 shows the morphology of nickel on graphene. The average nanocluster size was around 123 nm (Figure 3a inset) with surface roughness ($R_a$) of 2.95 nm. The height of the nanoclusters ranged between 20 to 50 nm. In contrast, nickel on $SiO_2$ is rather uniform after deposition, with a mean average surface roughness of 0.46 nm. The dramatic difference can be explained by the surface free energy difference between graphene and $SiO_2$. The surface free energy of graphene is on the order of 46.7 mJ/m$^2$, which is 500 times smaller than nickel particles (2.45 J/m$^2$) [30]. This suggests that metals will have sufficient mobility to migrate on graphene surface until reach to a high energy boundaries such as defective sites, resulting in cluster nucleation. In our case, the relative large cohesive energy (4.44 eV) for nickel favors the growth of the nanocluster in 3 D coverage, giving large clusters. On the other hand, $SiO_2$ has surface energy on the order of 4.32 J/m$^2$. Nickel tend to uniformly nucleate onto the $SiO_2$, resulting in continuous films. While the size of the resulting nickel clusters are comparable to metals grown on transferred graphene [32], it is much larger and more randomly distributed than most literature reported cluster growth on graphene in the form of moiré patterns [19-23]. This could be attributed to the amorphous $SiO_2$ substrate and other extrinsic factors (such as minor residue leftover) during graphene transfer process. Nonetheless, the resulting coalescent clusters show charge transfer with the graphene scaffold, which is evidenced by Raman spectroscopy (Figure 4). Consistent with recent literature [31], the fwhm (full width half maximum) of the G band is broadened and the corresponding peak position is downshifted after nickel deposition. This is presumably due to the weakening of the graphene C-C bonds by the hybridization effect between the nickel d orbital and graphene p orbital at the interface. This could be further evidenced by the downshift of the 2D band shown in Figure 4b.

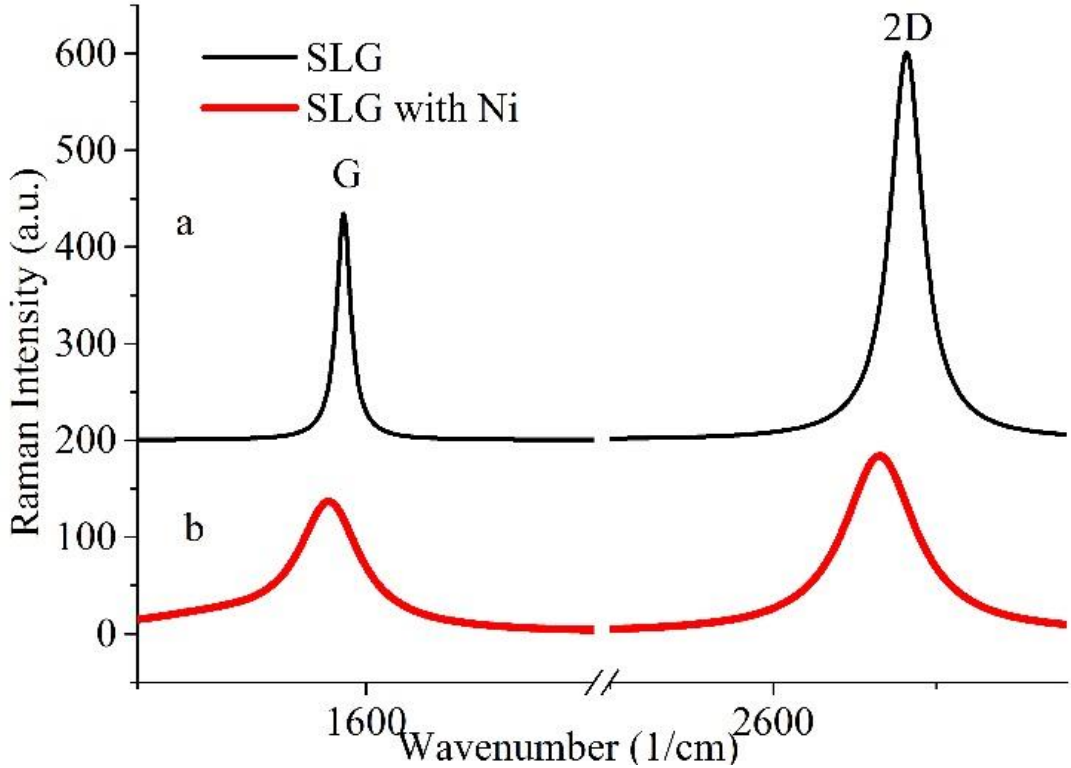

Figure 4. (Color online) Raman spectra of graphene: a) as-transferred, b) decorated with nickel cluster.

To further understand the graphene hybridized nickel-$NiO_x$ system, exchange bias measurements were used to detect the changes before and after high vacuum annealing. Figure 5 shows the hysteresis loops (25 $^0$C) before and after high vacuum annealing at 600 $^0$C. Before high vacuum annealing, the hysteresis loop exhibits horizontal shifts, a strong indication that antiferromagnetic nickel oxide is pinning the ferromagnetic moment from metallic nickel. After the sample is annealed in high vacuum at 600 $^0$C for 1 hr, the asymmetry of the hysteresis loop is significantly reduced. This is consistent with a significant amount of reduction of nickel oxide as observed from the XPS analysis. With little presence of oxide, the exchange bias effect is dominated by nickel nanoclusters. Interestingly, it is worth noting that there are increases not only in zero field remanence for the annealed sample, but also in the coercive magnetization field for the same sample. The slightly increase in zero field remanence can be attributed to the reduction of antiferromagnetic nickel oxide, resulting in weight gain of ferromagnetic nickel. A similar observation was reported for nickel/nickel oxide on multilayer graphene nanosheet, in which ferromagnetic nickel showed high zero field remanence [33]. There are multiple factors that might contribute to the change in the coercivity, including particle size, shape, crystallinity and exchange anisotropy [34]. The impact of which one is dominant in a specific system is controversial and not well understood [35][36]. According to an idealized Meiklejohn-Bean theory, coercivity ($H_c$) is inversely proportional to the saturation magnification field ($M_s$) and proportional to the volume anisotropy constant ($K_F$) for the ferromagnetic materials [35]. With a small change in $M_s$ in our specific case, the increase in $H_c$ is related to $K_F$ change. The details of which factors dictate the observed phenomenon will be a subject of future study. Nonetheless, it is worth mention that a similar $H_c$ increase was seen on a control sample (Ni/SiO$_2$) with a much small magnitude.

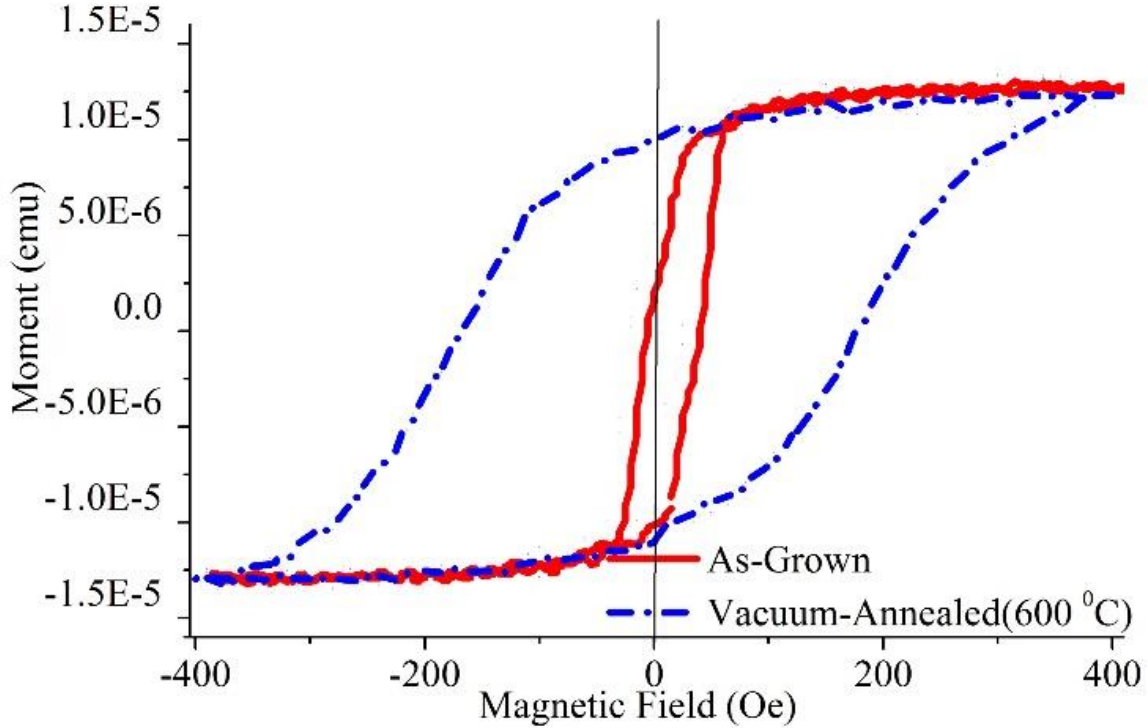

Figure 5. (Color online) Hysteresis loops (300 K) of nickel clusters: (1) as-grown (solid curve), (2) 600 $^{0}$C annealed (dotted curve).

## 4. Conclusions

In summary, we demonstrate a simple method to hybridize randomly oriented nickel clusters with CVD graphene on $SiO_2$ substrate. The post fabrication oxidation effect of the nickel clusters was investigated and an alternative pathway to effectively reduce oxide using high vacuum annealing is suggested. The vacuum annealed nickel cluster is proved to be stable after air exposure. The findings provide a pathway to assemble metal cluster on high quality CVD graphene on an arbitrary substrate. The nickel clusters on electronically conductive graphene after vacuum annealing showed improved coercivity, which can be further explored as a building blocks for functional magnetic composite materials [37].

`

## Acknowledgments

The authors acknowledge helpful discussion with Professor Carl Ventrice and the support from College of Nanoscience and Engineering at SUNY Polytechnic Institute.